\documentclass[a4paper,12pt,twoside]{article}
\usepackage{a4wide}
\usepackage[centertags]{amsmath}
\usepackage{amsfonts}
\numberwithin{equation}{section}

\let\rho\varrho
\let\phi\varphi
\newcommand{\pE}{\text{E}  }     % E
\newcommand{\mR}{\mathbb{R}}

\newcommand{\mZ}{\mathbb{Z}}

\newcommand{\Dd}{\text{d}}      % Diferentations "d"
\newcommand{\DD}{\text{D}}      % Gross "D"
\newcommand{\Dp}{\partial}      % Partielle Differentation
\newcommand{\Di}{\Dd}           % Integrations "d"
\DeclareMathOperator{\Tr}{Tr}
\DeclareMathOperator{\tr}{tr}

\newcommand{\qed}{\ \hfill $\Box$\par}

\newcommand\DDS{\text{\hbox to 0pt{/\hss}D}}
\newcommand\DpS{\text{\hbox to 0pt{/\hss}}\partial}

\begin{document}

\begin{titlepage}

\noindent\hfill TUM-HEP-289/97

%\noindent\hfill\today

\vfill

\begin{center}
  {\Huge Brane Tensions and Coupling Constants\\
    from within M-Theory}

  \vspace{1.5cm}

  Jan O. Conrad\footnote{E-mail: Jan.Conrad@physik.tu-muenchen.de}

  \smallskip

  {\it Institut f\"ur Theoretische Physik\\
    Physik Department\\
    Technische Universit\"at M\"unchen\\
    D-85747 Garching, FRG}

  \vfill

  {\bf Abstract}
  
  \bigskip

  \parbox{0.9\textwidth}{%
    Reviewing the cancellation of local anomalies of M-theory on
    $\mR^{10} \times S^1/\mZ_2$ the Yang-Mills coupling constant on
    the boundaries is rederived. The result is $\lambda^2 = 2^{1/3}
    (2\pi) (4\pi \kappa^2)^{2/3}$ corresponding to $\eta =
    \lambda^6/\kappa^4 = 256 \pi^5$ in the ``upstairs'' units used by
    Ho\v{r}ava and Witten and differs from their calculation. It is
    shown that these values are compatible with the standard membrane
    and fivebrane tensions derived from the M-theory bulk action. In
    view of these results it is argued that the natural units for
    M-theory on $\mR^{10} \times S^1/\mZ_2$ are the ``downstairs''
    units where the brane tensions take their standard form and the
    Yang-Mills coupling constant is $\lambda^2 = 4\pi (4\pi
    \kappa^2)^{2/3}$.}

\end{center}

\vfill

\end{titlepage}

\section{Introduction and summary}

This paper is devoted to the recalculation of coupling constants and
brane tensions of M-theory on $\mR^{10} \times S^1/\mZ_2$
\cite{HWI,HWII} by purely M-theoretic methods thereby clarifying the role
of the ``upstairs'' and ``downstairs'' approaches.

To begin, in the conventions of Ho\v{r}ava and Witten\cite{HWII}, the
bulk action of M-theory on $\mR^{10} \times S^1/\mZ_2$ in the
``upstairs'' approach is given by
\begin{equation}\label{ISMU}
  S_M =
  \frac{1}{2\Bar{\kappa}^2} \int_{M^{11}_U} \Di^{11} x
  \sqrt{-g} \: \Bigl( R + \ldots \Bigr.
\end{equation}
where ``upstairs'' refers to the fact that $M^{11}_U$ is defined to be
$\mR^{10} \times S^1$, all fields are $\mZ_2$ symmetric and
$\mZ_2$ is generated by $x^{11} \mapsto - x^{11}$ (for the details see
\cite{HWII}).

In the ``downstairs'' approach one works on the manifold $M^{11} =
\mR^{10} \times S^1/\mZ_2 = M^{11}_U / \mZ_2$ which, by modding out
$M^{11}_U$ with $\mZ_2$, aquires two ten-dimensional boundaries
denoted by $M^{10}$ and ${M'}^{10}$. The action in this approach is
related to \eqref{ISMU} by application of the $\mZ_2$ symmetry giving
\begin{equation}\label{ISM}
  S_M =
  \frac{1}{\Bar{\kappa}^2} \int_{M^{11}} \Di^{11} x
  \sqrt{-g} \: \Bigl( R + \ldots \Bigr.
\end{equation}
It should be noted that \eqref{ISMU} is not an action on $M^{11}_U$
simply because the degrees of freedom live on $M^{11}$, not on
$M^{11}_U$. In this respect \eqref{ISMU} is just a rewriting of
\eqref{ISM} convenient to carry out calculations.

Now the action written down by an eleven-dimensional observer sitting
in the bulk is
\begin{equation}\label{ISMII}
  S_M =
  \frac{1}{2\kappa^2} \int \Di^{11} x
  \sqrt{-g} \: \Bigl( R + \ldots \Bigr.
\end{equation}
where $\kappa$ was chosen as in \cite{DLM} and it is {\it this} action
which was used to derive membrane and fivebrane tensions in the bulk
\cite{DLM,AlwisI}. It is, however, physically reasonable to demand
that bulk brane tensions are independent of whether any boundaries
exist arbitrarily far away from the observer or whether a dimension is
compact at arbitrary large scales. Were that not the case we would get
different brane tensions in the limit of decompactifying M-theory on
$\mR^{10} \times S^1$ and M-theory on $\mR^{10} \times S^1/\mZ_2$.
Thus ``M-theory'' would be a rather strange construct. Of course we do
not know, a priori, whether M-theory is a physically reasonable theory
and one has to rely on detailed calculations. In order to do so, one
therefore has check against brane tensions derived from \eqref{ISM} or
\eqref{ISMU} after converting to the units used in \eqref{ISMII}.
Comparison of \eqref{ISMII} and \eqref{ISM} gives the conversion
relation
\begin{equation}\label{kappa}
  \Bar{\kappa}^2 = 2 \kappa^2
\end{equation}

Now, as shown in \cite{HWI, HWII}, the $\mZ_2$-symmetry mods out half
of the supersymmeries and requires the theory to be supplemented by
$\pE_8$-Yang-Mills-multiplets living on $M^{10}$ and ${M'}^{10}$.
Recalculation of the coupling constant $\lambda$ of these multiplets in
section~\ref{local} using anomaly cancellation on the boundaries will
lead to
\begin{equation}
  \lambda^2 = 2^{1/3} (2\pi) (4\pi\Bar{\kappa}^2)^{2/3}
  = 4\pi (4\pi \kappa^2)^{2/3}
\end{equation}
and
\begin{equation*}
  \eta = \frac{\lambda^6}{\kappa^4} = (4\pi)^5 \qquad
  \Bar{\eta} = \frac{\lambda^6}{\Bar{\kappa}^4} = 256 \pi^5
\end{equation*}
which clearly stands in contradiction to \cite{HWII} and
\cite{AlwisII}. However, by the same calculation one can also
determine the fivebrane tension $T_5$ itself and, by fixing the
normalization of the four form field strength $K_4$ as in
\cite{AlwisII}, the membrane tension $T_2$ (see section
\ref{membrane}):
\begin{equation}
  T_2 = \left(\frac{(2\pi)^2}{2 \kappa^2} \right)^{1/3} =
        \left(\frac{(2\pi)^2}{\Bar{\kappa}^2} \right)^{1/3}
  \qquad\qquad
  T_5 = \left(\frac{2\pi}{(2 \kappa^2)^2} \right)^{1/3} =
        \left(\frac{2\pi}{\Bar{\kappa}^4} \right)^{1/3}
\end{equation}
These tensions are in perfect agreement with those derived in
\cite{AlwisI}, provided one uses the units of \eqref{ISMII} as
explained above. The tensions obey the interrelations
\begin{equation}\label{IDuality}
  2 \kappa^2 T_2 T_5 = \Bar{\kappa}^2 T_2 T_5 = 2 \pi \qquad\qquad
  \frac{T_5}{T_2^2} = \frac{1}{2\pi}    
\end{equation}
which show that, in the ``upstairs'' units, the membrane/fivebrane
duality relation has not its natural form $2\Bar{\kappa}^2 T_2 T_5 = 2
\pi$ which one would naively assume when looking at the integral
\eqref{ISMU}. This is due to the fact that \eqref{ISMU} is not an
action on $M^{11}_U$ but on $M^{11}$.

From that viewpoint it is therefore more natural to use the units of
the ``downstairs'' approach. The bosonic bulk action including the
``fivebrane term'' (see section \ref{not}) is then\footnote{When going
  ``upstairs'' the $2\pi/2\kappa^2 T_5$-term becomes
  $2\pi/2\Bar{\kappa}^2 T_5$ and by application of \eqref{IDuality}
\begin{equation*}
  \frac{1}{2} T_2 \int_{M^{11}}
    C_3 \wedge \frac{1}{24 (2\pi)^4}
    \left( \frac{1}{8} \tr R^4 -
      \frac{1}{32}(\tr R^2)^2 \right)
\end{equation*}}
\begin{equation}\label{ISMIII}
  \begin{aligned}
    S_M =\,& \frac{1}{2\kappa^2} \int_{M^{11}}
    R \Omega - \frac{1}{2} K_4 \wedge *K_4
    + \frac{1}{6} C_3 \wedge K_4 \wedge K_4 \\
    & + \frac{2\pi}{2\kappa^2 T_5} \int_{M^{11}}
    C_3 \wedge \frac{1}{24 (2\pi)^4}
    \left( \frac{1}{8} \tr R^4 -
      \frac{1}{32}(\tr R^2)^2 \right)
  \end{aligned}
\end{equation}
together with the Super-Yang-Mills actions on $M^{10}$ and ${M'}^{10}$
\begin{equation}\label{IYMaction}
  \begin{gathered}
    S_{YM} = - \frac{1}{\lambda^2} \int_{M^{10}} \Di^{10} x \sqrt{g}
    \tr\left( \frac{1}{4}F_{AB} F^{AB} \right) \\
    S'_{YM} = - \frac{1}{\lambda^2} \int_{{M'}^{10}} \Di^{10} x \sqrt{g}
    \tr\left( \frac{1}{4}F'_{AB} {F'}^{AB} \right)
  \end{gathered}
\end{equation}
where fermionic fields were suppressed. In that connection the bulk
action \eqref{ISMIII} directly corresponds to the bulk action of
M-theory on $\mR^{10} \times S^1$ and so do the brane tensions.

The outline of the rest of the paper is as follows. In
section~\ref{not} notations and conventions are introduced.
Section~\ref{local} covers the calculation of the Yang-Mills coupling
constant and the fivebrane tension reviewing anomaly cancellation on
one boundary (M-theory on $\mR^{11}/\mZ_2$) on the lines of
\cite{HWII, AlwisII}.  In section~\ref{membrane} the membrane tension
is derived from the results of section~\ref{local} as in
\cite{AlwisII, Flux}.

\section{Notations and conventions}
\label{not}

In this and the next section we will consider only one boundary, that
is M-theory on $M^{11} = \mR^{10} \times \mR/\mZ_2 = \mR^{10} \times
\mR^{+}$. Therefore we have $M^{11}_U = \mR^{10} \times \mR$. The
bosonic part of the supergravity action used in \cite{HWII} in the
``upstairs'' approach is\footnote{Our conventions are mainly as in
  \cite{HWI,HWII}: Compared to those used in \cite{HWII} we have $K_4
  = \sqrt{2}G^{HW}$, $C_3 = 6\sqrt{2} C^{HW}$ and $R = - R^{HW}$.
  $\Omega$ denotes the volume measure $\sqrt{-g} \: \Dd x^1 \wedge
  \ldots \wedge \Dd x^{11}$. When using $1=\Gamma^1 \ldots
  \Gamma^{11}$ the positive sign of the $CKK$-term is forced upon us
  by supersymmetry. This appears when checking the terms of the form
  $\Bar{\eta}\psi K^2$ containing nine gamma matrices in the
  supersymmetry variation of the action (see \cite{CJS}).}
\begin{equation}\label{SM}
  \begin{aligned}
    S_M =\,& \frac{1}{2\Bar{\kappa}^2} \int_{M^{11}_U}
    R \Omega - \frac{1}{2} K_4 \wedge *K_4
    + \frac{1}{6} C_3 \wedge K_4 \wedge K_4 \\
    & + \frac{2\pi}{2\Bar{\kappa}^2 T_5} \int_{M^{11}_U}
    C_3 \wedge \frac{1}{24 (2\pi)^4}
    \left( \frac{1}{8} \tr R^4 -
      \frac{1}{32}(\tr R^2)^2 \right)
  \end{aligned}
\end{equation}
where the last term, which subsequently will be called fivebrane term,
is required i.e.\ by anomaly cancellation on the fivebrane\cite{DLM}.

The fields are supposed to be invariant under the $\mZ_2$-symmetry
acting by $x^{11} \mapsto - x^{11}$ thus introducing an orbifold
singularity at $x^{11} = 0$. $M^{10}$ will denote the locus of these
points endowed with the orientation $\Dd x^1 \wedge \ldots \wedge \Dd
x^{10}$.  At $M^{10}$ the theory is supplemented by an
$\pE_8$-Super-Yang-Mills multiplet of positive chirality Majorana-Weyl
fermions. The bosonic part of the action for these fields is (with
units as in \cite{HWII})
\begin{equation}\label{YMaction}
  S_{YM} = - \frac{1}{\lambda^2} \int_{M^{10}} \Di^{10} x \sqrt{g}
  \tr\left( \frac{1}{4}F_{AB} F^{AB} \right)
\end{equation}
where $\tr = 1/30 \Tr$ and $\Tr$ denotes the trace in the adjoint
representation of $\pE_8$. Uppercase indices from the beginning of the
alphabet run from 1 to 10.

As shown in \cite{HWII} local supersymmetry requires the
Bianchi-identity of $K_4$ be modified\footnote{ In \cite{HWII} this
  was given by
  \begin{equation*}
    \Dd G^{HW}_{11ABCD} = - 3 \sqrt{2} \frac{\Bar{\kappa}^2}{\lambda^2}
    \delta(x^{11}) \frac{1}{24} \tr F_{AB} F_{CD} + \text{cyclic
    permutations of $ABCD$} 
  \end{equation*}
  Using $K_4 = \sqrt{2} G^{HW}$ together with the relation
  \begin{equation*}
    (\tr F^2)_{ABCD} = \frac{1}{4} \tr F_{AB} F_{CD} + \text{cyclic
      permutations of $ABCD$} 
  \end{equation*}
  yields \eqref{modBianchi}.
}
\begin{equation}\label{modBianchi}
  \Dd K_4 = - \frac{\Bar{\kappa}^2}{\lambda^2}
  \tr F^2 \, \delta(x^{11}) \Dd x^{11}
\end{equation}

To simplify the discussion of anomalies and especially the Wess-Zumino
consistency condition hidden in the descend equations it is useful to
introduce a BRST-like operator $s$. It generalizes gauge
transformations and has the following properties
\begin{equation} \label{BRST}
  \begin{aligned}
    s A &= - \Dd \Lambda - [A,\Lambda] = -\DD \Lambda \qquad
        & s F &= [ F,\Lambda ] \qquad
        & s \Lambda &= - \frac{1}{2} [\Lambda,\Lambda] \\
    s \omega &= - \Dd \Theta - [\omega,\Theta] = -\DD \Theta \qquad
        & s R &= [ R, \Theta ] \qquad
        & s \Theta &= - \frac{1}{2} [\Theta,\Theta] \\
    s^2 &= 0 \qquad & s\Dd + \Dd s &= 0
        & s \int &= \int s
  \end{aligned}
\end{equation}
Using $s$ one can write down the following forms
\begin{equation}
  \begin{aligned}
    \omega_{4L} &= \tr R^2 \qquad & \omega_{4Y} &= \tr F^2 \\
    \omega_{3L} &= \tr (\omega R - \tfrac{1}{3} \omega^3) \qquad
    & \omega_{3Y} &= \tr (AF - \tfrac{1}{3} A^3) \\
    \omega^1_{2L} &= \Tr (\Theta \Dd \omega) &
    \omega^1_{2Y} &= \Tr (\Lambda \Dd A)
  \end{aligned}
\end{equation}
obeying the equations
\begin{equation}
  \omega_4 = \Dd \omega_3 \qquad s\omega_4 = 0
\end{equation}
and the first two of the so called {\it descend equations} (see i.e.\ 
\cite{AGG} and references therein; the explicit expression for
$\omega^2_1$ will not be used)
\begin{equation}\label{descend}
  \begin{aligned}
    s &\omega_3   &\,+\,& \Dd \omega^1_2 &\,= 0 \\
    s &\omega^1_2 &\,+\,& \Dd \omega^2_1 &\,= 0
  \end{aligned}
\end{equation}

\section{Review of local anomaly cancellation}
\label{local}

The starting point is the modified Bianchi-identity in the ``upstairs''
approach
\begin{equation}
  \Dd K_4 = c \omega_4 \delta(x^{11}) \: \Dd x^{11}
\end{equation}
(for the moment let $\omega_4$ be defined as $-\omega_{4Y}$;
\eqref{modBianchi} then gives $c=\Bar{\kappa}^2/\lambda^2$).
Demanding the definition of $K_4$ be $K_4 = \Dd C_3$ {\it outside}
of $M^{10}$ leads to
\begin{equation}\label{modK4}
  K_4 = \Dd C_3 + c \omega_3 \delta(x^{11}) \: \Dd x^{11}
\end{equation}
which in addition is motivated by the modification of the three-form
field strength known from ten-dimensional Super-Yang-Mills-theory
\cite{SUYM10D}.

As, by the equations of motion, $K_4$ may not contain delta functions,
$C_3$ must contain a step funtion (where $\epsilon$ is defined as an
odd function obeying $\epsilon'(x^{11}) = 2 \delta(x^{11})$)
\begin{equation}\label{modC3}
  C_3 = \frac{1}{2} \epsilon(x^{11}) c \omega_3 + \Hat{C}_3
\end{equation}
and $\Hat{C}_3$ contains no step functions supported at $M^{10}$.
Inserting this into \eqref{modK4} gives
\begin{equation}\label{modK4II}
  K_4 = \frac{1}{2} c \epsilon(x^{11}) \omega_4 + \Dd \Hat{C}_3
\end{equation}

Applying the $s$ operator to \eqref{modC3} gives the transformation
law of $C_3$
\begin{equation}\label{sC3}
  sC_3
  = \frac{1}{2} \epsilon(x^{11}) \: cs\omega_3 + s\Hat{C}_3
  = - \frac{1}{2} c \epsilon(x^{11}) \: \Dd\omega^1_2 + s \Hat{C}_3
\end{equation}
Now $sK_4=0$, that is gauge invariance of $K_4$, yields by \eqref{modK4}
and \eqref{sC3}
\begin{equation}
  s\Hat{C}_3 = - \Dd \xi
\end{equation}
where the sign is convention, of course. However, $\Hat{C}_3$ is
perfectly regular at $M^{10}$ and so is $\xi$. In addition $\xi$ is odd
under parity because $C_3$ is odd under parity and so, by modding out
with $\mZ_2$, we have
\begin{equation}
  \xi_{AB} (x^{11}) = - \xi_{AB} (-x^{11})
\end{equation}
Therefore $\xi$ vanishes when pulled back to $M^{10}$.

In the ``downstairs'' approach \eqref{modC3}, \eqref{modK4II} and
\eqref{sC3} are now given by taking the limit $x^{11} \mapsto 0$ from
$x^{11} > 0$
\begin{equation}\label{downstairs}
  \begin{aligned}
    C_3 &= \frac{1}{2} c \omega_3 + \Hat{C}_3 \\
    K_4 &= \frac{1}{2} c \omega_4 + \Dd\Hat{C}_3 \\
    sC_3 &= -\frac{1}{2} c \Dd \omega^1_2 - \Dd\xi
  \end{aligned}
\end{equation}

Using these relations one can now calculate the variation of the $CKK$
term under gauge transformations
\begin{equation}\label{sSupstairs}
  sS_{CKK} = \frac{1}{2\Bar{\kappa}^2} \int_{M^{11}_U}
  \frac{1}{6} \: sC_3 \wedge K_4 \wedge K_4
\end{equation}
This is performed easiest going to the ``downstairs''
approach\footnote{%
  The calculation in the ``upstairs'' approach is a little tedious but
  gives the same result. One especially encounters integrals over
  $K^2_4 \: \delta(x^{11}) \: \Dd x^{11}$ the relevant part of which
  is, of course,
  \begin{equation*}
    \int_{-\infty}^{+\infty} \epsilon^2(x) \delta(x) \Di x
    = \int_{-\infty}^{+\infty}
      \frac{1}{6} \frac{\Dd \epsilon^3(x)}{\Dd x} \Di x
    = \frac{1}{3}
  \end{equation*}} using \eqref{downstairs}.
Then \eqref{sSupstairs} gives\footnote{%
  There are, however, some subtleties in this calculation. First, by
  the definition of the operator $s$ and the form $w^1_2$ Stokes law
  receives a minus sign, because $\Lambda$ and $\Theta$ are one-forms
  of the exterior algebra of $s$ which anticommutes with $\Dd$ and $s$
  commutes with integration as given in \eqref{BRST}.  Second, this
  sign is cancelled by the fact that $M^{10}$ has the opposite
  orientation compared to the induced orientation on $\Dp M^{11}$
  which is $-\Dd x^1 \wedge \ldots \wedge \Dd x^{10}$.}
\begin{equation}\label{sSCKK}
  \begin{aligned}
    sS_{CKK}
    &= \frac{1}{\Bar{\kappa}^2} \int_{M^{11}}
       \frac{1}{6} \: sC_3 \wedge K_4 \wedge K_4 \\
    &= \frac{1}{\Bar{\kappa}^2} \int_{M^{11}}
       - \frac{1}{6} \: \left( \frac{1}{2}c \: \Dd \omega^1_2 + \Dd \xi \right)
       \wedge K_4 \wedge K_4 \\
    &= \frac{1}{\Bar{\kappa}^2} \int_{M^{11}}
       - \frac{1}{6} \:\Dd \left(
         \left( \frac{1}{2}c \omega^1_2 + \xi \right)
         \wedge K_4 \wedge K_4 \right) \\
    &= \frac{1}{\Bar{\kappa}^2} \int_{\Dp M^{11} = -M^{10}}
       \frac{1}{6\cdot 2} \: c \: \omega^1_2 \wedge K_4 \wedge K_4 \\
    &= \frac{1}{2\Bar{\kappa}^2} \int_{M^{10}}
       - \frac{1}{6} \: c \: \omega^1_2 \wedge K_4 \wedge K_4 \\
    &= \frac{1}{2\Bar{\kappa}^2} \int_{M^{10}}
       - \frac{1}{6} \: c^3 \: \omega^1_2 \wedge \frac{w^2_4}{4}
  \end{aligned}
\end{equation}

To check cancellation of purely non-gravitational anomalies (setting
$\Theta=0$) resulting from the Super-Yang-Mills multiplet
\eqref{YMaction} one has to compute the anomaly polynomial for
ten-dimensional positive chirality Majorana-Weyl fermions in the
adjoint representation of $\pE_8$. This is given by (see i.e.\ 
\cite{HWII,Flux,AGG,GSW})
\begin{equation}
  \begin{aligned}
    Q_{12} &= \frac{1}{2} \Tr \frac{1}{6!}
              \left( \frac{i}{2\pi} F \right)^6
     = \frac{1}{2} \frac{30}{6!} \tr \left( \frac{i}{2\pi} F \right)^6
     = -\frac{1}{2} \frac{30}{8 \cdot 6!} (\tr F^2)^3 \\
    &= \frac{1}{2 \cdot 48}\frac{1}{(2\pi)^6} \:
       \omega_4 \wedge \frac{\omega^2_4}{4}
  \end{aligned}  
\end{equation}
where the well known relation
\begin{equation}
  \tr F^6 = \frac{30^2}{7200} (\tr F^2)^3
  = \frac{1}{8} (\tr F^2)^3
\end{equation}
valid for $\pE_8$ was used.

The anomaly is then
\begin{equation}
  s\Gamma = 2\pi \int_{M^{10}} Q^1_{10}
\end{equation}
where $Q^1_{10}$ is determined by $Q_{12} = \Dd Q_{11}$ and the
descend equations (see i.e.\ \cite{AGG})
\begin{equation}
  \begin{aligned}    
    s &Q_{11}   &\,+\,& \Dd Q^1_{10} &\,= 0 \\
    s &Q^1_{10} &\,+\,& \Dd Q^2_{9}  &\,= 0
  \end{aligned}
\end{equation}
giving\footnote{%
  When using the covariant anomaly one gets $Q^1_{10} \sim 6 \Tr
  (\Lambda F^5)$ from $Q_{12} \sim \Tr F^6$ by transgression and the
  descend equations. This eventually yields
  \begin{equation*}
    Q^1_{10} = - 6 \times \frac{1}{2 \cdot 48} \frac{1}{(2\pi)^6} \:
    \tr (\Lambda F) \wedge \frac{\omega^2_4}{4}
  \end{equation*}
  Therefore the anomaly in \cite{HWII} was a factor of 6 too big
  compared to \eqref{Q110}, which was partially cancelled by a factor
  of 3 introduced from (3.1) to (3.2) of \cite{HWII} leaving an
  uncancelled factor of 2.}
\begin{equation}\label{Q110}
  Q^1_{10} = \frac{1}{2 \cdot 48} \frac{1}{(2\pi)^6} \: \omega^1_2 \wedge
  \frac{\omega^2_4}{4}
\end{equation}

Anomaly cancellation is then determined via \eqref{sSCKK}
\begin{equation}\label{cancel}
  0 = s\Gamma + sS = s\Gamma + sS_{CKK}
  = \left( \frac{1}{2 \cdot 48}\frac{1}{(2\pi)^5}
           - \frac{1}{6 \cdot 2} \frac{c^3}{\Bar{\kappa}^2} \right)
    \int_{M^{10}} \omega^1_2 \wedge \frac{w^2_4}{4}
\end{equation}
where the variation from the fivebrane term, which can not be
cancelled by purely non-gravitational anomalies due to factors of $R$,
has been omitted.  One then gets
\begin{equation}\label{coverkappa}
  \frac{c^3}{\Bar{\kappa}^2}
  = \frac{\Bar{\kappa}^4}{\lambda^6}
  = \frac{1}{8 (2\pi)^5}
\end{equation}
Rewriting this using $\Bar{\kappa}^2 = 2 \kappa^2$ we get
\begin{equation}
  \frac{\kappa^4}{\lambda^6}
  = \frac{1}{(4\pi)^5}
\end{equation}

As discussed in length in \cite{HWI,HWII} taking gravitational and mixed
anomalies into account the anomaly polynomial gets modified to
\begin{equation}\label{Q12}
  Q_{12} = \frac{1}{2 \cdot 48} \frac{1}{(2\pi)^6}
  \: \omega_4 \wedge \left(
    \frac{\omega^2_4}{4} +
    \frac{1}{8} \tr R^4 - \frac{1}{32}(\tr R^2)^2
  \right)
\end{equation}
where $\omega_4$ is now defined by
\begin{equation}
  \omega_4 = \frac{1}{2} \tr R^2 - \tr F^2
  = \frac{1}{2} \omega_{4L} - \omega_{4Y}
\end{equation}
Now the variation of the action stemming from the fivebrane term must
be included
\begin{equation}\label{sS5brane}
  \begin{aligned}
    sS_{\text{\scriptsize 5-brane}}
    &= \frac{2\pi}{2\Bar{\kappa}^2 T_5} \int_{M^{11}_U}
       sC_3 \wedge \frac{1}{24 (2\pi)^4}
       \left( \frac{1}{8} \tr R^4 -
         \frac{1}{32}(\tr R^2)^2 \right) \\
    &= -\frac{2\pi}{2\Bar{\kappa}^2 T_5} \frac{c}{24 (2\pi)^4}
       \int_{M^{10}}
       \omega^1_2
       \left( \frac{1}{8} \tr R^4 -
         \frac{1}{32}(\tr R^2)^2 \right)
  \end{aligned}
\end{equation}
(compare to \eqref{sSCKK}).  Taking the modification of \eqref{Q12} in
\eqref{Q110} into account\footnote{It should be noted that $Q^1_{10}$
  is not uniquely determined by the descend equations when taking
  gravitational and mixed anomalies into account. However this does
  play no role in the cancellation mechanism but only leads to some
  ambiguity in a local counterterm not mentioned here. For the
  details, see \cite{GSW}, chapter 13.5.3..} anomaly cancellation gives
\begin{equation}
  \begin{aligned}
    0 &= s\Gamma + sS
    = s\Gamma + sS_{CKK} + sS_{\text{\scriptsize 5-brane}} \\
    &= \left( \frac{1}{2 \cdot 48} \frac{1}{(2\pi)^5}
              - \frac{2\pi}{2\Bar{\kappa}^2 T_5} \frac{c}{24 (2\pi)^4}
            \right)
       \int_{M^{10}}
       \omega^1_2
       \left( \frac{1}{8} \tr R^4 -
         \frac{1}{32}(\tr R^2)^2 \right)
  \end{aligned}
\end{equation}
Therefore the fivebrane tension is
\begin{equation}\label{T5}
  T_5^3
  = \frac{2\pi}{\Bar{\kappa}^4}
  = \frac{2\pi}{(2 \kappa^2)^2}
\end{equation}

\section{The membrane tension}
\label{membrane}

The results of the last section can be used to derive the macroscopic
membrane tension by the observation of \cite{AlwisII} and \cite{Flux}
that the normalization of $K_4$ on the boundary is related to global
anomalies on the worldvolume of macroscopic membranes in the bulk.
This is expressed by the equation \cite{AlwisII,Flux}\footnote{In
  \cite{Flux} $\pm T_2 K_4$ was {\it by definition} denoted by $G$.}
\begin{equation}
  \left. \pm \frac{T_2 K_4}{2\pi} \right|_{M^{10}}
  = \frac{1}{16\pi^2}
    \left( \frac{1}{2} \tr R^2 - \tr F^2 \right)
\end{equation}
This is to be compared to \eqref{downstairs}
\begin{equation}
  \left. K_4 \right|_{M^{10}}
  = \frac{1}{2} c \omega_4
  = \frac{1}{2} c \left( \frac{1}{2} \tr R^2 - \tr F^2 \right)
\end{equation}
which eventually yields by \eqref{coverkappa} ($c$ is positive for the
chirality chosen in section~\ref{not})
\begin{equation}
  T_2 = \frac{1}{4\pi c}
  = \left(\frac{(2\pi)^2}{2 \kappa^2} \right)^{1/3}
  = \left(\frac{(2\pi)^2}{\Bar{\kappa}^2} \right)^{1/3}
\end{equation}

\section*{Acknowledgements}

I would like to thank H.-P. Nilles, D. Matalliotakis, A. Kehagias, M.
Olechowski, M. Yamaguchi and M. Zucker for many discussions. I am also
indebted to S.P. de Alwis and E. Witten. This work was supported by
the European Commission programs ERBFMRX-CT96-0045 and CT96-0090 and
by a grant from Deutsche Forschungsgemeinschaft SFB-375-95.

\end{document}